\documentclass[aps,prl,twocolumn,superscriptaddress]{revtex4}
\usepackage{graphics}
\usepackage{epsfig}
\usepackage{amsmath}
\usepackage{amssymb}
\begin{document}
\newcommand{\hrho}{\widehat{\rho}}
\newcommand{\homega}{\widehat{\omega}}
\newcommand{\hI}{\widehat{I}}
\newcommand*{\spr}[2]{\langle #1 | #2 \rangle}
\newcommand*{\bbN}{\mathbb{N}}
\newcommand*{\bbR}{\mathbb{R}}
\newcommand*{\cB}{\mathcal{B}}
\newcommand*{\E}{\cal{E}}
\newcommand{\supp}{\rm{supp}}
\newcommand*{\eps}{\varepsilon}
\newcommand*{\id}{I}
\newcommand{\orho}{\overline{\rho}}
\newcommand{\omu}{\overline{\mu}}
\newcommand*{\half}{{\frac{1}{2}}}
\newcommand*{\ket}[1]{| #1 \rangle}
\newcommand{\trho}{{\widetilde{\rho}}_n^\gamma}
\newcommand*{\bra}[1]{\langle #1 |}
\newcommand*{\proj}[1]{\ket{#1}\bra{#1}}
\newcommand{\otrho}{{\widetilde{\rho}}_n^{\gamma 0}}
\newcommand{\hsigma}{{\widehat{\sigma}}}
\newcommand{\tsigma}{{\widetilde{\sigma}}}
\newcommand{\tpi}{{\widetilde{\pi}}}
\newcommand{\tlambda}{{{\Lambda}}}
\newcommand{\tM}{{\widetilde{M}}}
\newcommand{\tomega}{{\widetilde{\omega}}}
\newcommand{\be}{\begin{equation}}
\newcommand{\bea}{\begin{eqnarray}}
\newcommand{\eea}{\end{eqnarray}}
\newcommand{\tr}{\mathrm{Tr}}
\newcommand*{\Hmin}{H_{\min}}
\newcommand{\rank}{\mathrm{rank}}
\newcommand{\tends}{\rightarrow}
\newcommand{\uS}{\underline{S}}
\newcommand{\oS}{\overline{S}}
\newcommand{\ee}{\end{equation}}
\newcommand{\ep}{\eps^{\prime}}
\newcommand{\n}{{(n)}}
\newtheorem{definition}{Definition}
\newtheorem{theorem}{Theorem}
\newtheorem{proposition}{Proposition}
\newtheorem{lemma}{Lemma}
\newtheorem{defn}{Definition}
\newtheorem{corollary}{Corollary}
\newcommand{\qed}{\hspace*{\fill}\rule{2.5mm}{2.5mm}}

\newenvironment{proof}{\noindent{\it Proof}\hspace*{1ex}}{\qed\medskip}
\def\reff#1{(\ref{#1})}
%


\title{Max- Relative Entropy of Entanglement, {\em{alias}} Log Robustness}


\author{Nilanjana Datta}
\email{N.Datta@statslab.cam.ac.uk}
\affiliation{Statistical Laboratory, DPMMS, University of Cambridge, Cambridge CB3 0WB, UK}

\date{\today}

\begin{abstract}
Properties of the {\em{max- relative entropy of entanglement}}, defined
in \cite{ND}, are investigated, and its significance as an upper bound to 
the one-shot rate for {\em{perfect}} entanglement dilution, under a 
particular class of quantum operations, is discussed.
It is shown that 
it is in fact equal to another known entanglement monotone, namely 
the {\em{log robustness}}, defined in \cite{BP}. It is known that 
the latter is not asymptotically continuous and it is not known whether
it is weakly additive. However, by suitably modifying the max- relative
entropy of entanglement we obtain a quantity which is seen to 
satisfy both these properties. In fact, the modified quantity is shown
to be equal to the regularised relative entropy of entanglement.
\end{abstract}

\pacs{03.65.Ud, 03.67.Hk, 89.70.+c}

\maketitle

\section{Introduction}
In \cite{renatophd}, Renner introduced two important entropic quantities, 
called the min- and max- entropies. Recently, the operational meanings of these 
quantities, i.e., their relevance with regard to actual information-processing
tasks, was elucidated in \cite{robert}. Further, two new relative entropy quantities, 
which act as parent quantities for these min- and max- entropies, were introduced in 
\cite{ND}. These were, namely, the max-relative entropy, $D_{\max}(\rho||\sigma)$, 
and the min-relative entropy, 
$D_{\min}(\rho||\sigma)$. Here $\rho$ denotes a state and
$\sigma$ denotes a positive operator. Various properties of these
quantities were proved in \cite{ND}. In particular, it was
shown that 
$$D_{\min}(\rho||\sigma)\le S(\rho||\sigma)\le D_{\max}(\rho||\sigma),$$
where $S(\rho||\sigma)$ is the relative entropy of $\rho$ and $\sigma$.

In addition, it was shown in \cite{ND} that the minimum over all separable
states, $\sigma$, of $D_{\max} (\rho||\sigma)$, defines a (full) entanglement
monotone \cite{full} for a bipartite state $\rho$. This quantity, referred to as the 
the {\em{max-relative entropy of entanglement}} and denoted by $E_{\max}(\rho)$,
was proven to be an upper bound to the relative entropy of entanglement, $E_R(\rho)$
\cite{VP}.

In this paper we investigate further properties of $E_{\max}(\rho)$ and discuss
its significance.
We prove that it is quasiconvex, i.e., for a mixture of states
${\rho = \sum_{i=1}^n p_i \rho_i}$,
$E_{\max}(\rho) \le \max_{1\le i\le n} E_{\max}(\rho_i).$ 
We also infer that it is not asymptotically continuous \cite{ascont},
and does not reduce
to the {\em{entropy of entanglement}} for pure bipartite states (that is, to the 
entropy of the reduced state of either of the two parties).  We do so 
by showing that $E_{\max}(\rho)$ is in fact equal to another known entanglement 
monotone, namely, the {\em{log robustness}} \cite{BP}: $LR_g(\rho) := \log (1 + R_g(\rho)).$ 
Here $R_g(\rho)$ denotes the global 
robustness \cite{harrow} of $\rho$, which is a measure of the amount of
noise that can be added to an entangled state $\rho$ before it 
becomes unentangled (separable). By suitably modifying 
$E_{\max}(\rho)$, we arrive at a quantity, which we denote by
$\E_{\max}(\rho)$, and which is asymptotically
continuous and weakly additive \cite{weak}.
Asymptotic continuity \reff{asc} is proved by showing that $\E_{\max}(\rho)$ 
is equal to the regularised 
relative entropy of entanglement $E_R^\infty(\rho)$ \cite{eisert}, 
for which this property has been proved \cite{matthias}.
The necessary modifications involve $(i)$ ``smoothing'' $E_{\max}(\rho)$ 
to obtain the {\em{smooth max-relative entropy of entanglement}}
$E_{\max}^\eps(\rho)$, for any fixed $\eps >0$. (This is similar to the smoothing
introduced by Renner \cite{renatophd} to obtain the smooth R\'enyi entropies from
the min- and max- entropies mentioned above); $(ii)$ regularising, and 
$(iii)$ taking the limit $\eps \rightarrow 0$ (see the following sections).

It would be natural to proceed analogously with the min-relative entropy and
define a quantity, $E_{\min}(\rho)$, to be the minimum over all separable
states, $\sigma$, of $D_{\min} (\rho||\sigma)$.
However, it can be shown \cite{violate} that $E_{\min}(\rho)$ is not a full
entanglement monotone. It {\em{can}} increase on average under local operations and classical comunication
(LOCC). 
Instead $E_{\min}(\rho)$ satisfies a weaker condition of monotonicity 
under LOCC maps, that is, 
$E_{\min}(\rho) \ge E_{\min}(\Lambda(\rho)),$ for any LOCC operation 
$\Lambda$. Nevertheless, as for the case of $E_{\max}(\rho)$, a ``smoothing'' 
of $E_{\min}(\rho)$, 
followed by regularisation,
yields a quantity which is equal to $E_R^\infty(\rho)$, in the limit of the 
smoothing parameter $\eps \rightarrow 0$. This will be presented in 
a forthcoming paper \cite{FBND}.

In \cite{BP} it was shown that $E_{R}^\infty(\rho)$ is equal to both 
the entanglement cost and the distillable entanglement under 
the set of quantum operations which do not generate any entanglement
asymptotically [for details, see \cite{BP}]. This gives an operational 
significance to the regularised version of the 
smooth max- and min- relative entropies of entanglement, 
in the limit $\eps \rightarrow 0$.

For a given $\eps >0$, the smoothed versions, $E^\eps_{\max}(\rho)$ 
and $E^\eps_{\min}(\rho)$, of the max- and min- relative entropies
of entanglement,
also have operational interpretations. They arise as optimal rates
of entanglement manipulation protocols involving separability-preserving
maps. A quantum operation $\Lambda$ is said to be a 
separability-preserving map if $\Lambda(\sigma)$ is separable 
for any separable state $\sigma$. These maps constitute 
the largest class of operations which cannot create entanglement and
contains the class of separable operations \cite{bennett, rains, VP}.
[See \cite{BP} for details]. The quantities $E^\eps_{\max}(\rho)$ 
and $E^\eps_{\min}(\rho)$ can 
be interpreted as one-shot rates of entanglement dilution and
distillation protocols involving separability-preserving maps, 
for a given bound
on the corresponding probabilities of error, i.e., when the
probability of error associated with the protocol
is bounded above by the smoothing parameter $\eps$. This is analogous to the
interpretation of the $\eps$-smooth R\'enyi entropies, 
as one-shot rates of various protocols
\cite{renatophd, wolf2, RenWol04b}, when the
probability of error is at most $\eps$.  Evaluation of these one-shot
rates for the entanglement manipulation protocols, 
will be presented in a forthcoming paper \cite{FBND}.

The max-relative entropy of entanglement (or log robustness), 
$E_{\max}(\rho)$, provides an upper bound to the one-shot {\em{perfect}} 
entanglement cost, not under LOCC maps, but under quantum operations 
which generate an entanglement (as measured by the global robustness) 
of at most $1/R_g(\rho)$. We shall refer to such maps as $\alpha_\rho$-separability preserving (or $\alpha_\rho$-SEPP) maps, with $\alpha_\rho =1/R_g(\rho)$. This is elaborated below. 

We start the main body of our paper with some mathematical preliminaries. Next
we recall the definitions of the relevant relative entropy quantities and 
entanglement 
monotones, and prove that $E_{\max}(\rho)$ is quasiconvex. We then show that it is equal to the global 
log robustness, and that it does {{not}} in general reduce to the relative entropy
of entanglement for pure states. Next we define the smooth max-relative entropy of entanglement and 
$\E_{\max}(\rho)$, and prove 
that the latter is weakly additive. Our main 
result is given in Theorem \ref{main1}, which states that ${\E}_{\max}(\rho)= E_R^\infty(\rho)$ \cite{point}.

\section{Mathematical Preliminaries}
\label{prelim}
Let ${\cal{B}}({\cal{H}})$ denote the algebra of linear operators
acting on a finite-dimensional Hilbert space ${\cal{H}}$. The von
Neumann entropy of a state $\rho$, i.e., a positive operator of unit
trace in ${\cal{B}}({\cal{H}})$, is given by $S(\rho) = - \tr \rho
\log \rho$. Throughout this paper, we take the logarithm to base $2$
and all Hilbert spaces considered are finite-dimensional. In fact,
since in this paper we consider bipartite states, the underlying 
Hilbert space is given by ${\cal{H}}= {\cal{H}}_A \otimes {\cal{H}}_B$. Let
${\cal{D}}$ denote the set of states in ${\cal{B}}({\cal{H}})$, and
let ${\cal{S}} \subset {\cal{D}}$ denote the set of separable states. 
Further, let ${\cal{S}}_n$ denote the set of separable states in
${\cal{B}}({\cal{H}}^{\otimes n})$.

The trace distance between two operators $A$ and $B$ is given by
\be
||A-B||_1 := \tr\bigl[\{A \ge B\}(A-B)\bigr] -
 \tr\bigl[\{A < B\}(A-B)\bigr]
\ee
The fidelity of states $\rho$ and $\rho'$ is defined to be
$$ F(\rho, \rho'):= \tr \sqrt{\rho^{\half} \rho' \rho^{\half}}.
$$
The trace distance between two states $\rho$ and $\rho'$ is
related to the fidelity $ F(\rho, \rho')$ as follows (see (9.110) of \cite{nielsen}):
\be
  \frac{1}{2} \| \rho - \rho' \|_1
\leq
  \sqrt{1-F(\rho, \rho')^2}
\leq
  \sqrt{2(1-F(\rho, \rho'))} \ .
\label{fidelity}
\ee

We also use the ``gentle measurement'' lemma \cite{winter99,ogawanagaoka02}:
\begin{lemma}\label{gm} For a state $\rho$ and operator $0\le \Lambda\le I$, if
$\mathrm{Tr}(\rho \Lambda) \ge 1 - \delta$, then
$$||\rho -   {\sqrt{\Lambda}}\rho{\sqrt{\Lambda}}||_1 \le {2\sqrt{\delta}}.$$
The same holds if $\rho$ is only a subnormalized density operator.
\end{lemma}

\section{Definitions of min- and max- relative entropies}
\label{non-smooth}

\begin{definition}
  The \emph{max- relative entropy} of a state
  $\rho$ and a positive operator $\sigma$ is given by
  \be
    D_{\max}(\rho|| \sigma)
  :=
    \log \min\{ \lambda: \, \rho\leq \lambda \sigma \}
  \label{dmax}
\ee
\end{definition}
Note that $D_{\max}(\rho|| \sigma)$ is well-defined if 
$\supp\, \rho \subseteq \supp\, \sigma$.

\begin{definition}
The \emph{min- relative entropy} of a state
  $\rho$ and a positive operator $\sigma$ is given by
  \be
    D_{\min}(\rho|| \sigma)
  :=    - \log \tr\bigl(\pi\sigma\bigr) \ ,
  \label{dmin}
\ee
  where $\pi$ denotes the projector onto $\supp\, \rho$, the support of $\rho$. It is well-defined if 
$\supp\, \rho$ has non-zero intersection with $\supp\, \sigma$.
\end{definition}

Various properties of $D_{\min}(\rho|| \sigma)$ and $D_{\max}(\rho|| \sigma)$ were proved
in \cite{ND}. In this paper we shall use the following properties of the max- relative entropy, $D_{\max}(\rho|| \sigma)$:
\begin{itemize}
\item{The max- relative entropy is monotonic under completely positive trace-preserving 
(CPTP) maps, i.e., for
a state $\rho$, a positive operator $\sigma$, and a CPTP map $\Lambda$:
\be D_{\max}(\Lambda(\rho)||\Lambda(\sigma))\le  D_{\max}(\rho||\sigma)
\label{mono}
\ee}
\item{The max- relative entropy is quasiconvex, i.e.,
for two mixtures of states, $\rho:=\sum_{i=1}^n p_i \rho_i$ and
$\omega:= \sum_{i=1}^n p_i \omega_i$, 
\be
D_{\max}(\rho||\omega) \le \max_{1\le i\le n} D_{\max}(\rho_i || \omega_i).
\label{quasid}
\ee}
\item{$D_{\max}(\rho\otimes \rho||\omega\otimes \omega) =2 D_{\max}(\rho||\omega) .$ This property
follows directly from the definition \reff{dmax}.}
\end{itemize}

The min- and max- (unconditional and conditional) entropies, introduced 
by Renner
in \cite{renatophd} are obtained from $D_{\min}(\rho|| \sigma)$ and $D_{\max}(\rho|| \sigma)$
by making suitable substitutions for the positive operator $\sigma$ (see \cite{ND} for details). 

\section{Smooth min- and max- relative entropies}
\label{smooth}
\emph{Smooth} min- and max- relative entropies are generalizations of the above-mentioned  
relative entropy measures, involving an additional \emph{smoothness} parameter
$\eps \geq 0$. For $\eps = 0$, they reduce to the
\emph{non-smooth} quantities. 
\begin{definition} \label{def:smoothentropies} For
  any $\eps \geq 0$, the \emph{$\eps$-smooth min-} and
  \emph{max-relative entropies} of a bipartite state $\rho$ relative to a
  state $\sigma$ are defined by
\[
    D_{\min}^{\eps}(\rho || \sigma)
  :=
    \max_{\bar{\rho} \in B^{\eps}(\rho)} D_{\min}(\bar{\rho} || \sigma)
  \]
  and
  \be
    D_{\max}^{\eps}( \rho|| \sigma )
  :=
    \min_{\bar{\rho} \in B^{\eps}(\rho)} D_{\max}(\bar{\rho} || \sigma)
  \label{epsmax}
\ee
  where $B^{\eps}(\rho) := \{\bar{\rho} \geq 0: \, \| \bar{\rho} - \rho
  \|_1 \leq \eps, \tr(\bar{\rho}) \leq \tr(\rho)\}$.
\end{definition}

The following two lemmas are used to prove our main result, Theorem \ref{main1}. 

\begin{lemma} \label{lem5n}
  Let $\rho_{A B}$ and  $\sigma_{AB}$ be density operators, let $\Delta_{A B}$ be a positive operator, and let $\lambda \in \bbR$  such that
   \[
    \rho_{A B} \leq 2^{\lambda} \cdot \sigma_{AB} + \Delta_{A B} \ .
  \]
  Then $D_{\max}^{\eps}(\rho_{A B}||\sigma_{AB}) \le \lambda$ for any $\eps \geq \sqrt{8 \tr(\Delta_{A B})}$.
\end{lemma}

\begin{lemma} \label{lem6n}
  Let $\rho_{A B}$ and $\sigma_{AB}$ be density operators. Then
  \[
    D_{\max}^{\eps}(\rho_{A B}|\sigma_B) \le \lambda
  \]
  for any $\lambda \in \bbR$ and
  \[
    \eps = \sqrt{8 \tr\bigl[\{\rho_{A B} > 2^{\lambda} \sigma_{AB} \} \rho_{A B} \bigr]} \ .
  \]
\end{lemma}

The proofs of these lemmas are analogous to the proofs of Lemmas 5 and 6 of \cite{smooth}
and are given in the Appendix for completeness.

\section{The max-relative entropy of entanglement}
\label{entm}
For a bipartite state $\rho$, the {max-relative entropy of entanglement}
\cite{ND} is given by
\be E_{\max}(\rho):= \min_{\sigma \in {\cal{S}}} D_{\max} (\rho||\sigma),
\label{entmeasure}
\ee
where the minimum is taken over the set  ${\cal{S}}$ 
of all separable states.

It was proved in \cite{ND} that
\be
E_{\max}(\rho) \ge E_R(\rho),
\ee
where $E_R(\rho) := \min_{\sigma \in {\cal{S}}} S (\rho||\sigma), $
the {\em{relative entropy of entanglement}} of the state $\rho$.

That $E_{\max}(\rho)$ is a full entanglement monotone follows from the fact that
$D_{\max}(\rho||\sigma)$ satisfies a set of sufficient criteria \cite{VP} 
which ensure that $E_{\max}(\rho)$ has the following properties: $(a)$ it 
vanishes if and only if $\rho$ is separable,
$(b)$ it is left invariant by local unitary operations and $(c)$ it does not
increase on average under LOCC. This was proved in \cite{ND}.

\begin{lemma} The {max-relative entropy of entanglement} $E_{\max}(\rho)$ 
is quasiconvex, i.e., 
for a mixture of states
${\rho = \sum_{i=1}^n p_i \rho_i}$,
\be
E_{\max}(\rho) \le \max_{1\le i\le n} E_{\max}(\rho_i).
\label{convex}
\ee
\end{lemma}
\begin{proof}
For each state $\rho$, let $\sigma_\rho$ be a separable state for which
$$
E_{\max}(\rho) = D_{\max}(\rho||\sigma_\rho ).
$$
Since the set of separable states is convex, and the max-relative entropy 
is quasiconvex \reff{quasid},
we have
\bea
E_{\max}\Bigl(\sum_i p_i \rho_i\Bigr)
&\le&D_{\max}\Bigl(\sum_i p_i\rho_i || \sum_i p_i \sigma_{\rho_i}\Bigr)\nonumber\\
&\le & \max_i D_{\max}\Bigl(\rho_i || \sigma_{\rho_i}\Bigr)\nonumber\\
&=& \max_i E_{\max}(\rho_i)
\eea
\end{proof}

Since $E_{\max}(\rho)$ is given by a minimisation over separable states, it
is subadditive. Let $\sigma$ be a separable state
for which $E_{\max}(\rho) = D_{\max}(\rho ||\sigma).$ Then, 
\bea
E_{\max}(\rho \otimes \rho)  &=& 
\min_{\omega \in {\cal{S}}_2}  D_{\max}(\rho \otimes \rho||\omega)\nonumber\\
&\le &   D_{\max}(\rho \otimes \rho||\sigma \otimes \sigma)\nonumber\\
&=& 2 D_{\max}(\rho||\sigma) = 2 E_{\max}(\rho).
\eea

\begin{lemma}
\label{logrobust}
The {max-relative entropy of entanglement} $E_{\max}(\rho)$ of a bipartite
state $\rho$ is equal to its {\em{global log robustness}} of
entanglement \cite{BP}, which is defined as follows:
\be
LR_g(\rho) := \log \bigl(1 + R_g(\rho)\bigr),
\ee
where $R_g(\rho)$ is the {\em{global robustness of
entanglement}}\cite{harrow}, given by
$$
R_g(\rho) = \min_{s \in \mathbb{R}} \Bigl\{s\ge 0 : \exists \,\omega \in {\cal{D}} \,\,{\rm{s.t.}}\,\, 
\frac{1}{1+s}\rho + \frac{s}{1+s}\omega \in  {\cal{S}} \Bigr\} 
$$
\end{lemma}
\begin{proof}
We can equivalently write  $R_g(\rho)$ as follows:
\bea 
R_g(\rho) &=& \min_{s \in \mathbb{R}} \Bigl\{s\ge 0 : \exists \,\omega \in {\cal{D}}  \,\,{\rm{s.t.}}\,\, 
\rho + s\omega = (1+s) \sigma, \sigma \in  {\cal{S}} \Bigr\}\nonumber\\
&=& \min_{s \in \mathbb{R}} \Bigl\{s\ge 0 : \exists \,\sigma \in {\cal{S}}  \,\,{\rm{s.t.}}\,\,  
\rho \le (1+s) \sigma \Bigr\},
\eea
since, defining $\tomega := (1+s)\sigma - \rho$, we see that 
$\tr \, \tomega = 1 + s - 1 = s$, hence allowing us to write $\tomega = s \omega$
for some $\omega \in  {\cal{D}}$.  
Hence, 
$$\log (1+ R_g(\rho)) = \min_{\sigma \in {\cal{S}}} D_{\max}( \rho|| \sigma).$$
\end{proof}
\begin{definition} A state $\pi$ for which $$\rho + R_g(\rho) \pi = (1 + R_g(\rho))\sigma,$$ for some separable state $\sigma$, is referred to as an optimal
state for $\rho$ in the global robustness of entanglement.
\end{definition}
\medskip

\noindent
It was shown in \cite{harrow} that for a pure bipartite 
state $\rho= |\psi\rangle\langle \psi| \in {\cal{B}}({\cal{H}}_A \otimes {\cal{H}}_B)$,
$$R_g(\rho) = \Bigl(\sum_{i=1}^m \lambda_i\Bigr)^2 - 1,$$
where the $\lambda_i$. $i=1, \ldots, m$, denote the Schmidt coefficients of $|\psi\rangle$. This implies
that for the pure state $\rho= |\psi\rangle\langle \psi|$, the max-relative entropy
of entanglement is given by
\be
E_{\max}(\rho) = \log (1 + R_g(\rho)) = 2 \log \Bigl( \sum_{i=1}^m \lambda_i\Bigr),
\label{pure}
\ee
i.e., twice the logarithm of the sum of the square roots of the eigenvalues 
of the reduced density matrix $\rho_\psi^A := \tr_B |\psi\rangle\langle \psi|$. Hence for a pure state $\rho$, $E_{\max}(\rho)$ does {\em{not}} in general 
reduce to its {\em{entropy of entanglement}} (i.e., the von Neumann entropy 
of $\rho_\psi^A$), even though it does so for a maximally entangled state.
Let $\Psi_M \in {\cal{B}}({\cal{H}_A}\otimes {\cal{H}_B})$ denote a maximally entangled state (MES) of rank $M$, i.e.,
$\Psi_M = |\Psi_M\rangle\langle\Psi_M|,$ with 
$$|\Psi_M\rangle = \frac{1}{\sqrt{M}} \sum_{i=1}^M |i\rangle|i\rangle.$$
Then, 
$$ E_{\max} (\Psi_M) = \log M = S(\tr_A \Psi_M).$$
Moreover, $R_g(\Psi_M) = M - 1.$

Note that the right hand side of \reff{pure} is equal to the expression 
for another known entanglement monotone, namely the 
{\em{logarithmic negativity}} \cite{martin}
$$LN(\rho) := \log ||\rho^\Gamma||_1,$$
for the pure state $\rho=|\psi\rangle\langle \psi|$. Here $\rho^\Gamma$ denotes
the partial transpose with respect to the subsystem $A$, and
$||\omega||_1 = \tr \sqrt{w^\dagger \omega}.$ It is known that $LN(\rho)$
is additive \cite{martin} for pure states, and we therefore have
\be
E_{\max}(|\psi\rangle \langle \psi| \otimes |\phi\rangle \langle \phi|)
= E_{\max}(|\psi\rangle \langle \psi|) + E_{\max}(|\phi\rangle \langle \phi|)
.
\label{ad1}
\ee
This additivity relation does not extend to mixed states in general. However,
the following relation can be proved to hold \cite{FBND}:
\bea
E_{\max}(\rho \otimes \Psi_M) 
&=& E_{\max}(\rho)+E_{\max}(\Psi_M)\nonumber\\
&=&E_{\max}(\rho)+ \log M.
\label{key}
\eea

As mentioned in the Introduction $E_{\max}(\rho)$ provides an 
upper bound to the 
one-shot {\em{perfect}} entanglement cost, under quantum
operations which generate an entanglement of at most $R_g(\rho)$. This is
elaborated below.

In entanglement dilution the aim is to obtain a state $\rho$ from a 
maximally entangled state. This cannot necessarily be done by using 
a single copy of the maximally entangled state and acting on it by a 
LOCC map. However,
a single perfect copy of $\rho$ {\em{can}} be obtained from a single copy of 
a maximally entangled state if one does not restrict the quantum operation 
employed to be a LOCC map but instead allows quantum operations which
generate an entanglement of at most $1/R_g(\rho)$. Before proving this, let us
state the definition of one-shot perfect entanglement cost of a state
under a quantum operation $\Lambda$. 

\begin{definition} A real number $R$ is said to be an {\em{achievable}} 
one-shot perfect dilution rate, for a state $\rho$, under a quantum 
operation $\Lambda$, if $\Lambda(\Psi_M) =\rho$ and $\log M \le R$.
\end{definition}

\begin{definition}
The one-shot perfect entanglement cost of a state
under a quantum operation $\Lambda$ is given by $E_{c,\lambda}^{(1)} = \inf R,$
where the infimum is taken over all achievable rates.
\end{definition}  

Consider the quantum operation $\Lambda$ which acts on any state $\omega$
as follows:
\be
\Lambda_M(\omega) = \tr( \Psi_M \omega) \rho + (1 - \tr( \Psi_M \omega))\pi,
\label{qop}
\ee
where $\pi$ is an optimal state for $\rho$ in the global robustness 
of entanglement. It was shown in \cite{BP} 
that for $M= 1 + s$, where $s = R_g(\rho)$, 
the quantum operation $\Lambda_M$ is an $(1/s)$-separability preserving map (SEPP),
i.e., for any separable state $\sigma$: 
$$R_g(\Lambda (\sigma)) \le 1/s.$$
In other words, the map $\Lambda$ as defined by \reff{qop}, is a quantum 
operation which generates an entanglement corresponding to a global robustness
of at most $1/R_g(\rho)$.

Now if $\omega =  \Psi_M$, then $\Lambda(\omega) = \rho$, and hence a 
perfect copy of $\rho$ is obtained from a single copy of the maximally entangled
state $ \Psi_M$. The associated rate, $R$, of the one-shot entanglement 
dilution
protocol corresponding to the map $\Lambda_M$ satisfies the bound \cite{loose}:
\be
R \le \log M = \log (1 +s) = E_{\max}(\rho).
\label{upbdd}
\ee

The log robustness, $LR_g(\rho)$, is not asymptotically continuous \cite{BP} and it is not known
whether it is weakly additive. 
However, as mentioned in the Introduction, by suitable modifying $E_{\max}(\rho)$ 
one can arrive at a quantity
which is both asymptotically continuous and weakly additive. The necessary modifications involve
$(i)$ ``smoothing'' $E_{\max}(\rho)$ to obtain the {\em{smooth max-relative entropy of entanglement}}
$E_{\max}^\eps(\rho)$, for any fixed $\eps >0$; $(ii)$ regularising, and $(iii)$ taking the
limit $\eps \rightarrow 0$, as described below.

\section{Smooth Max-Relative entropy of entanglement and ${\E}_{\max}(\rho)$}

For any $\eps >0$, we define the {\em{smooth max-relative entropy of 
entanglement}}
of a bipartite state $\rho$, as follows:
\bea
{E}_{\max}^\eps (\rho) &:=& \min_{\bar{\rho} \in B^{\eps}(\rho)} E_{\max}(\bar{\rho})
\nonumber\\
&=& \min_{\bar{\rho} \in B^{\eps}(\rho)} \min_{\sigma\in {\cal{S}}} D_{\max}(\bar{\rho}||\sigma),
\nonumber\\
&=& \min_{\sigma\in {\cal{S}}} D_{\max}^\eps(\rho||\sigma),
\label{eqmax}
\eea
where $D_{\max}^\eps(\rho||\sigma)$ is the smooth max-relative entropy
defined by \reff{epsmax}. Further, we define its regularised version
\be
{\E}_{\max}^\eps (\rho):= \limsup_{n\rightarrow \infty}\frac{1}{n} E_{\max}^\eps(\rho^{\otimes n}),
\ee
and the quantity
\be
{\E}_{\max} (\rho) := \lim_{\eps \rightarrow 0} {\E}_{\max}^\eps (\rho)
\label{deff}
\ee

\begin{lemma} The quantity ${\E}_{\max} (\rho)$ characterizing a 
bipartite state $\rho  \in {\cal{B}}({\cal{H}})$ and defined by \reff{deff},
satisfies the following properties:
\begin{enumerate}
\item{It is weakly additive, i.e., for any positive integer $m$, 
\be
{\E}_{\max} (\rho^{\otimes m}) = m \, {\E}_{\max}(\rho).
\label{weak}
\ee}
\item{It is asymptotically continuous, i.e., for a given $\eps >0$, 
if $\rho_m \in {\cal{B}}({\cal{H}}^{\otimes m})$
is an operator for which $||\rho_m - \rho^{\otimes m} ||_1 \le \eps$, then
\be
\bigl|\frac{{\E_{\max}}(\rho_m) - {\E_{\max}}(\rho^{\otimes m})}{m}\bigr| \le f(\eps),
\label{asc}\ee
where $f(\eps)$ is a real function of $\eps$ such that $f(\eps)\rightarrow 0$
as $\eps \rightarrow 0$.}
\end{enumerate}
\end{lemma}
\begin{proof} Here we give the proof of $1$, by showing that 
${\E}_{\max} (\rho\otimes \rho) = 2{\E}_{\max}(\rho)$.
The proof of $2$ follows 
from Theorem \ref{main1} below since the regularized relative entropy of entanglement
$E_R^\infty(\rho)$, (defined by \reff{rel1}), is known to be asymptotically continuous \cite{matthias}.

We first prove that 
\be
{\E_{\max}}(\rho \otimes \rho) \le 2 {\E}_{\max}(\rho)
\label{part1}
\ee
Fix $\eps >0$. Then,
\bea
{E}_{\max}^\eps(\rho^{\otimes n})&=& \min_{\sigma_n \in {\cal{S}}_n} D^\eps_{\max}(\rho^{\otimes n}|| \sigma_n),\nonumber\\
&=&  \min_{\sigma_n \in {\cal{S}}_n} \min_{ \orho_n \in B^\eps(\rho^{\otimes n})} D_{\max}(\orho_{n}|| \sigma_n)
\label{opt}\\
&=&  D_{\max}(\rho_n^\eps|| \sigma_n^\eps),
\label{opt2}
\eea
where $\rho_n^\eps \in  B^\eps(\rho^{\otimes n})$ and $\sigma_n^\eps \in  {\cal{S}}_n$ are operators for which the minima in 
\reff{opt} are achieved.

Since $\rho_n^\eps \in  B^\eps(\rho^{\otimes n})$, we have that $|| \rho_n^\eps - \rho^{\otimes n}||_1 \le \eps,$ which
in turn implies that 
$$|| \rho_n^\eps\otimes \rho_n^\eps - \rho^{\otimes 2n}||_1 \le 2\eps.$$
Therefore, $ \rho_n^\eps\otimes \rho_n^\eps \in B^\eps(\rho^{\otimes 2n})$. 
Further, since $ \sigma_n^\eps\otimes \sigma_n^\eps \in {\cal{S}}_{2n}$,
we have
\bea
E_{\max}^{2 \eps} (\rho^{\otimes 2n}) &=& \min_{ \orho_{2n} 
\in B^\eps(\rho^{\otimes 2n})} \min_{\sigma_{2n} \in {\cal{S}}_{2n}}  D_{\max}(\orho_{2n}|| \sigma_{2n})\nonumber\\
&\le &  D_{\max}(\rho_n^\eps\otimes \rho_n^\eps||  \sigma_n^\eps \otimes  \sigma_n^\eps)\nonumber\\
&=& 2 D_{\max}(\rho_n^\eps||  \sigma_n^\eps)\nonumber\\
&=& 2 {E}_{\max}^\eps(\rho^{\otimes n}).
\eea
Hence, 
\be
\limsup_{n\rightarrow \infty}\frac{1}{n}E_{\max}^{2 \eps} \bigl((\rho\otimes\rho)^{\otimes  n}\bigr) \le 2 
\limsup_{n\rightarrow \infty}\frac{1}{n}E_{\max}^{\eps} (\rho^{\otimes n}),
\ee
that is,
${\E}_{\max}^{2\eps}(\rho \otimes \rho) \le 2 {\E}_{\max}^{\eps}(\rho)$. Taking the limit $\eps \rightarrow 0$ on either
side of this inequality yields the desired bound \reff{part1}.

In fact, the identity holds in \reff{part1}. This is simply because
\bea
{\E}_{\max}^{\eps}(\rho \otimes \rho)&=& \limsup_{n\rightarrow \infty}\frac{1}{n}E_{\max}^{\eps} 
((\rho\otimes \rho)^{\otimes n}),\nonumber\\
&=& 2\,\limsup_{n\rightarrow \infty}\frac{1}{2n}E_{\max}^{\eps} 
((\rho)^{\otimes 2n}),\nonumber\\
&=& 2\,{\E}_{\max}^{\eps}(\rho)
\label{29}
\eea
The last line of \reff{29} is proved \cite{fernando} by employing the monotonicity \reff{mono} of the max- relative
entropy under partial trace. We know that 
\bea
{\E}_{\max}^{\eps}(\rho)&:=& \limsup_{n\rightarrow \infty}\frac{1}{n}E_{\max}^{\eps} 
((\rho)^{\otimes n}),\nonumber\\
&\ge & \,\limsup_{n\rightarrow \infty}\frac{1}{2n}E_{\max}^{\eps} 
((\rho)^{\otimes 2n}),\label{id}
\eea
However, it can be proven that the identity always holds in 
\reff{id}. This is done by assuming that  
\be
{\E}_{\max}^{\eps}(\rho) >  \,\limsup_{n\rightarrow \infty}\frac{1}{2n}E_{\max}^{\eps} 
((\rho)^{\otimes 2n}),
\label{assume}
\ee
and showing that this leads to a contradiction.

The assumption \reff{assume} implies that there exists a sequence
of odd integers $n_i$ for which 
\be 
\limsup_{n_i \rightarrow \infty} \frac{1}{n_i} E_{\max}^\eps(\rho^{\otimes n_i}) 
>  
\limsup_{n_i  \rightarrow \infty} \frac{1}{n_i + 1} E_{\max}^\eps(\rho^{\otimes n_i + 1}). 
\ee 
Let $\rho_{n_i + 1} \in {\cal{H}}^{\otimes n_i + 1}$ be 
an operator in $B^\eps(\rho^{\otimes n_i + 1})$ for which
$E_{\max}^\eps(\rho^{\otimes n_i + 1}) = E_{\max}(\rho_{n_i + 1})$.  Then, using 
the monotonicity \reff{mono} of the max-relative entropy under partial trace, 
we have 
\bea 
 E_{\max}^\eps(\rho^{\otimes n_i + 1}) &=&E_{\max}(\rho_{n_i + 1}) \nonumber\\
&\ge& E_{\max}( \tr_{\cal{H}}(\rho_{n _i + 1}))\nonumber\\
&\ge& E_{\max}^\eps(\rho^{\otimes n_i}),
\eea 
since $\tr_{\cal{H}}(\rho_{n _i + 1}) \in B^\eps(\rho^{\otimes n_i})$. Therefore, 
\bea 
\limsup_{n_i \rightarrow \infty} \frac{1}{n_i} E_{\max}^\eps(\rho^{\otimes n_i}) &>& 
\limsup_{n_i  \rightarrow \infty} \frac{1}{n_i + 1} E_{\max}^\eps(\rho^{\otimes n_i + 1})\nonumber\\
&\ge & 
\limsup_{n_i \rightarrow \infty} \frac{1}{n_i + 1} E_{\max}^\eps(\rho^{\otimes n_i})\nonumber\\
&=& \limsup_{n_i \rightarrow \infty} \frac{1}{n_i} E_{\max}^\eps(\rho^{\otimes n_i}), 
\eea 
which is a contradiction.

\end{proof}
\section{Main Result}
Our main result is given by the following theorem.
\begin{theorem}
\label{main1}
\be
{\E}_{\max} (\rho)= E_R^\infty(\rho),
\ee
where, $E_R^\infty(\rho)$ denotes the regularized relative entropy of entanglement:
\be
E_R^\infty(\rho):= \lim_{n\rightarrow \infty} \frac{1}{n}E_R(\rho^{\otimes n}),
\label{rel1}
\ee
where, $E_R(\rho) = \min_{\sigma\in {\cal{S}}} S(\rho|| \sigma)$ is the relative
entropy of entanglement of $\rho$.
\end{theorem}
\begin{proof}
We first prove that $E_R^\infty(\rho) \le {\E}_{\max} (\rho)$.

Fix $\eps >0$.
\be
E_{\max}^\eps(\rho^{\otimes n}) = \min_{\sigma_n \in {{\cal{S}}}_n} D_{\max}^\eps (\rho^{\otimes n}||\sigma_n),
\ee
where ${{\cal{S}}}_n$ denotes the set of separable states in ${\cal{B}}({\cal{H}}^{\otimes n})$. In the above,
\be
D_{\max}^\eps (\rho^{\otimes n}||\sigma_n)= \min_{\orho_n \in B^\eps(\rho^{\otimes n})} 
D_{\max}(\orho_n||\sigma_n)
\label{eq1}
\ee
Let $\rho_n^\eps \in B^\eps(\rho^{\otimes n})$ be the operator for which the minimum is achieved 
in \reff{eq1}. Hence,
\be
E_{\max}^\eps(\rho^{\otimes n}) = \min_{\sigma_n \in {{\cal{S}}}_n}D_{\max} (\rho_n^\eps||\sigma_n)
\label{eq2}
\ee
Further, let $\tsigma_n$ be the separable state for which the minimum is achieved 
in \reff{eq2}. Hence,
\be
E_{\max}^\eps(\rho^{\otimes n}) = D_{\max} (\rho_n^\eps||\tsigma_n)
\label{eq3}
\ee
Since, 
$$D_{\max} (\rho_n^\eps||\tsigma_n) = \min\{\alpha : \rho_n^\eps \le 2^\alpha \tsigma_n\},$$
we have, 
\be
\rho_n^\eps \le 2^{E_{\max}^\eps(\rho^{\otimes n})}\tsigma_n.
\label{eq4}
\ee
Using \reff{eq4} and the operator monotonicity of the logarithm, we infer that
\be
S(\rho_n^\eps || \tsigma_n) \le E_{\max}^\eps(\rho^{\otimes n}),
\label{eq5}
\ee
since $\tr \rho_n^\eps \le \tr \rho^{\otimes n} = 1$.

From \reff{eq5} it follows that 
\be
\limsup_{n \rightarrow \infty} \frac{1}{n} S(\rho_n^\eps || \tsigma_n) \le {\E}_{\max}^\eps(\rho),
\label{eq6}
\ee
and hence, 
\be \limsup_{n \rightarrow \infty} \frac{1}{n} E_R(\rho_n^\eps) \le {\E}_{\max}^\eps(\rho),
\label{eq7}
\ee
where $E_R(\rho_n^\eps):= \min_{\sigma_n \in {\cal{S}}_n}  S(\rho_n^\eps || \sigma_n).$ 
It is known that $E_R(\rho)$ is asymptotically continuous. Hence,
\be
\frac{ E_R(\rho_n^\eps)}{n} \ge  \frac{E_R(\rho^{\otimes n})}{n} - f(\eps),
\label{eq8}
\ee
where $f(\eps)$ is a real function of $\eps$ satisfying $f(\eps) \rightarrow 0$ as $\eps \rightarrow 0$.
From \reff{eq7} and \reff{eq8} we obtain
$$ \limsup_{n \rightarrow \infty} \frac{1}{n} E_R(\rho^{\otimes n}) - f(\eps)\le {\E}_{\max}^\eps(\rho).$$
Taking the limit $\eps \rightarrow 0$ on both sides of the above inequality yields the desired bound:
$$ E_R^\infty (\rho)\le {\E}_{\max}(\rho).$$
\medskip

\noindent
We next prove the inequality $ E_R^\infty (\rho) \ge {\E}_{\max}(\rho).$

Consider the sequences $\hrho=\{\rho^{\otimes n}\}_{n=1}^\infty$ and 
$\hsigma=\{\sigma_\rho^{\otimes n}\}_{n=1}^\infty$, where 
$\sigma_\rho$ is the separable state for which
\be
E_R(\rho) = S(\rho||\sigma_\rho) \equiv \min_{\sigma'} S(\rho||\sigma').
\ee
For these two sequences, one can define the following quantity
\be
\overline{D}(\hrho \| \hsigma) 
:= \inf \Big\{ \gamma : \limsup_{n\rightarrow \infty} \mathrm{Tr}\big[ \{
\rho^{\otimes n} \ge 2^{n\gamma}\sigma_\rho^{\otimes n}\}\rho^{\otimes n} \bigr]=0\Bigr\}
\nonumber\\
\label{upd}
\ee
It s referred to as the {\em{sup-spectral divergence rate}} and arises in the
so-called Information Spectrum Approach \cite{nagaoka02, bd1}.
The Quantum Stein's Lemma (\cite{ogawa00} or equivalently {\em{Theorem 2}} of \cite{nagaoka02}) tells us that
\be
\overline{D}(\hrho \| \hsigma) = S(\rho||\sigma_\rho)
\ee

Let us choose 
\be
\lambda = \overline{D}(\hrho \| \hsigma) + \delta 
= E_R(\rho) + \delta,
\ee 
for some arbitrary $\delta >0$. It then follows from the definition \reff{upd} that
$$
\limsup_{n\rightarrow \infty} \mathrm{Tr}\big[ \{
\rho^{\otimes n} \ge 2^{n\lambda}\sigma_\rho^{\otimes n}\}\rho^{\otimes n} \Bigr]= 0
$$
In particular, for any $\eps > 0$ there exists $n_0 \in \bbN$, such that for all $n \geq n_0$.
\be
  \tr\bigl[\{\rho^{\otimes n} > 2^{n \lambda} \sigma_\rho^{\otimes n}\} \rho^{\otimes n} \bigr] < \frac{\eps^2}{8} \ .
\ee
Using Lemma \ref{lem6n} we infer that for all $n \ge n_0$, 
$$
D_{\max}^{\eps}(\rho^{\otimes n}||\sigma_\rho^{\otimes n}) \le n\lambda
= nE_R(\rho) + n\delta
$$
Hence, $E_{\max}^\eps(\rho^{\otimes n}) \le  nE_R(\rho) + n\delta$, and 
$$
{\E}_{\max}^\eps (\rho)\le E_R(\rho) + \delta.
$$
Moreover, since the above bound holds for any arbitrary $\delta>0$, we deduce that
${\E}_{\max}^\eps (\rho)\le E_R(\rho)$. Finally , taking the 
limit $\eps \rightarrow 0$ on both sides of this inequality yields
\be
{\E}_{\max} (\rho)\le E_R(\rho).
\ee
Using the weak additivity \reff{weak} of ${\E}_{\max} (\rho)$,
we obtain
\bea
\frac{1}{n}E_R(\rho^{\otimes n})&\ge & \frac{1}{n}{\E}_{\max}(\rho^{\otimes n})
\nonumber\\
&=& {\E}_{\max}(\rho).
\label{ll}
\eea
Taking the limit $n\rightarrow \infty$, on
both sides of \reff{ll}, yields
the desired bound 
$$ E_R^\infty(\rho) \ge {\E}_{\max}(\rho).$$
\end{proof}

\section{Appendix}
\label{lempfs} 
\noindent
{\bf{Proof of Lemma \ref{lem5n}}}

\begin{proof} Define
  \begin{align*}
  \alpha_{A B} & := 2^{\lambda} \cdot \sigma_{AB} \\
  \beta_{A B} & := 2^{\lambda} \cdot \sigma_{AB} + \Delta_{A B} \ .
  \end{align*}
 and
  \[
    T_{A B} := \alpha_{A B}^\half \beta_{A B}^{-\half} \ .
  \]
  Let $\ket{\Psi} = \ket{\Psi}_{A B R}$ be a purification of $\rho_{A B}$ and let $\ket{\Psi'} := T_{A B} \otimes \id_R \ket{\Psi}$ and $\rho'_{A B} := \tr_R(\proj{\Psi'})$.

Note that
\begin{align*}
  \rho'_{A B}
& =
  T_{A B} \rho_{A B} T_{A B}^{\dagger} \\
& \leq
  T_{A B} \beta_{A B} T_{A B}^{\dagger} \\
& =
  \alpha_{A B}
=
  2^{\lambda} \cdot \sigma_{AB} \ ,
\end{align*}
which implies $D_{\max}(\rho'_{A B}|\sigma_{AB}) \le \lambda$. It thus remains to be shown that
\begin{equation} \label{eq:distbound}
  \| \rho_{A B} - \rho'_{A B} \|_1
\leq
  \sqrt{8 \tr(\Delta_{A B})}
 \ .
\end{equation}

  We first show that the Hermitian operator
  \[
    \bar{T}_{A B} := \frac{1}{2} (T_{A B} + T_{A B}^\dagger) \ .
  \]
  satisfies
  \begin{equation} \label{eq:Tleqid}
    \bar{T}_{A B} \leq \id_{A B} \ .
 \end{equation}
 For any vector $\ket{\phi} = \ket{\phi}_{A B}$,
 \begin{align*}
   \| T_{A B} \ket{\phi} \|^2
 & =
   \bra{\phi} T_{A B}^\dagger T_{A B} \ket{\phi}
=
   \bra{\phi} \beta_{A B}^{-\half} \alpha_{A B} \beta_{A B}^{-\half} \ket{\phi} \\
 & \leq
   \bra{\phi} \beta_{A B}^{-\half} \beta_{A B} \beta_{A B}^{-\half} \ket{\phi}
 =
   \| \ket{\phi} \|^2
  \end{align*}
 where the inequality follows from $\alpha_{A B} \leq \beta_{A B}$.
 Similarly,
 \begin{align*}
   \| T_{A B} ^\dagger \ket{\phi} \|^2
 & =
   \bra{\phi} T_{A B} T_{A B}^\dagger \ket{\phi}
 =
   \bra{\phi} \alpha_{A B}^\half \beta_{A B}^{-1} \alpha_{A B}^{\half} \ket{\phi} \\
 & \leq
   \bra{\phi} \alpha_{A B}^{\half} \alpha_{A B}^{-1} \alpha_{A B}^{\half} \ket{\phi}
 =
   \| \ket{\phi} \|^2
  \end{align*}
 where the inequality follows from the fact that $\beta_{A B}^{-1} \leq \alpha_{A B}^{-1}$ which holds because the function $\tau \mapsto -\tau^{-1}$ is operator monotone on $(0, \infty)$ (see Proposition V.1.6 of \cite{bhatia}). We conclude that for any vector $\ket{\phi}$,
 \begin{align*}
  \| \bar{T}_{A B} \ket{\phi} \|
& \leq
   \frac{1}{2} \| T_{A B} \ket{\phi} + T_{A B}^{\dagger} \ket{\phi} \| \\
 & \leq
   \frac{1}{2} \| T_{A B} \ket{\phi} \| + \frac{1}{2} \| T_{A B}^{\dagger} \ket{\phi} \|
 \leq
   \| \ket{\phi} \| \ ,
 \end{align*}
 which implies~\eqref{eq:Tleqid}.

  We now determine the overlap between $\ket{\Psi}$ and $\ket{\Psi'}$,  \begin{align*}
    \spr{\Psi}{\Psi'}
   & =
    \bra{\Psi}  T_{A B} \otimes \id_R \ket{\Psi} \\
   & =
    \tr(\proj{\Psi} T_{A B} \otimes \id_R)
   =
    \tr(\rho_{A B} T_{A B}) \ .
 \end{align*}
 Because $\rho_{A B}$ has trace one, we have
\begin{align*}
    1 - |\spr{\Psi}{\Psi'}|
  & \leq
    1- \Re \spr{\Psi}{\Psi'}
  =
    \tr\bigl(\rho_{A B} (\id_{A B} - \bar{T}_{A B}) \bigr) \\
   & \leq
     \tr\bigl(\beta_{A B}  (\id_{A B} - \bar{T}_{A B})\bigr) \\
   & =
     \tr(\beta_{A B}) - \tr(\alpha_{A B}^{\half} \beta_{A B}^{\half}) \\
   & \leq
     \tr(\beta_{A B}) - \tr(\alpha_{A B})
   =
     \tr(\Delta_{A B}) \ .
  \end{align*}
  Here, the second inequality follows from the fact that, because of~\eqref{eq:Tleqid}, the operator $\id_{AB} - \bar{T}_{A B}$ is positive, and $\rho_{A B} \leq \beta_{A B}$. The last inequality holds because $\alpha_{A B}^{\half} \leq \beta_{A B}^{\half}$, which is a consequence of the operator monotonicity of the square root (Proposition V.1.8 of \cite{bhatia}).

Using \reff{fidelity} and the fact that the fidelity between two pure states is given by their overlap, we find
\begin{align*}
  \| \proj{\Psi} - \proj{\Psi'} \|_1
& \leq
  2 \sqrt{2(1-| \spr{\Psi}{\Psi'} |)} \\
& \leq
  2 \sqrt{2 \tr(\Delta_{A B})}
\leq
  \eps \ .
\end{align*}
Inequality~\eqref{eq:distbound} then follows because the trace distance can only decrease when taking the partial trace.
\end{proof}

\noindent
{\bf{Proof of Lemma \ref{lem6n}}}

\begin{proof}
  Let $\Delta^+_{A B}$ and $\Delta^-_{A B}$ be mutually orthogonal positive operators such that
  \[
    \Delta^+_{A B} - \Delta^-_{A B} = \rho_{A B} - 2^{\lambda} \sigma_{AB} \ .
  \]
  Furthermore, let $P_{A B}$ be the projector onto the support of $\Delta^+_{A B}$, i.e.,
  \[
    P_{A B} = \{\rho_{A B} > 2^{\lambda} \sigma_{AB} \} \ .
  \]
  We then have
  \begin{align*}
    P_{A B} \rho_{A B}  P_{A B}
  & =
    P_{A B} (\Delta^+_{A B} + 2^{\lambda} \sigma_{AB} - \Delta^-_{A B}) P_{A B} \\
  & \ge 
    \Delta^{+}_{A B}
  \end{align*}
  and, hence,
  \[
    \sqrt{8 \tr(\Delta^{+}_{A B})}
  \le
    \sqrt{8 \tr(P_{A B} \rho_{AB})\bigr)} = \eps \ .
  \]
  The assertion now follows from Lemma~\ref{lem5n} because
  \[
    \rho_{A B} \leq 2^{\lambda} \sigma_{AB} + \Delta^+_{A B} \ .
  \]
\end{proof}

\section*{Acknowledgements} The author is very grateful to Fernando Brandao for
invaluable discussions. She would also like to thank Milan Mosonyi and Yurii Suhov for helpful comments and Ismail Akhalwaya for carefully reading the paper.  


\begin{thebibliography}{1}
\bibitem{weak} An entanglement monotone $E(\rho)$ is weakly additive
if $E(\rho^{\otimes n}) = nE(\rho)$ for any positive integer $n$.
\bibitem{point} We would like to point out that in \cite{BP} a different modification of the log robustness was shown to
equal $E_R^\infty(\rho)$, by a method different to the one employed in this
paper. 
The main difference in the modification was that in \cite{BP}, the smoothing
parameter depended on the number $n$ of copies of the given state $\rho$.
\bibitem{loose} Note, however, that this bound is loose for bipartite states 
which are close to separable states. For an operational significance of the
robustness in relation to entanglement activation see \cite{brand}.
\end{thebibliography}
\end{document}